\documentclass[conference]{IEEEtran}
\usepackage{cite}
\usepackage{graphicx}
\usepackage[cmex10]{amsmath}
\usepackage{amssymb}
\usepackage{mathtools}
\usepackage{multirow}
\usepackage{notoccite}
\usepackage{color} 
\usepackage{epsfig}
\usepackage{latexsym}
\usepackage{pstricks}
\usepackage{pdfpages}
\usepackage{epstopdf}
\DeclareGraphicsExtensions{.eps}
\usepackage{algorithmic}
\usepackage{algorithmic}
\usepackage{algorithm}
\usepackage{multirow}
\usepackage{graphicx}

\twocolumn
\def\ninept{\def\baselinestretch{.95}\let\normalsize\small\normalsize}

\begin{document}
	\title{Double-RIS Communication with DF Relaying for Coverage Extension: Is One Relay Enough?}
	\author{\IEEEauthorblockN{Zaid Abdullah, Steven Kisseleff, Konstantinos Ntontin, Wallace Alves Martins,\\ Symeon Chatzinotas, and Bj$\ddot{\text{o}}$rn Ottersten \\
			\\ Interdisciplinary Centre for Security, Reliability and Trust (SnT), University of Luxembourg, Luxembourg. \\
			E-mails: \{zaid.abdullah,  steven.kisseleff, kostantinos.ntontin, wallace.alvesmartins,\\ symeon.chatzinotas,  bjorn.ottersten\}@uni.lu} }
	
	\maketitle
	\begin{abstract}
		In this work, we investigate the decode-and-forward (DF) relay-aided double reconfigurable intelligent surface (RIS)-assisted networks, where the signal is subject to reflections from two RISs before reaching the destination. Different relay-aided network architectures are considered for maximum achievable rate under a total power constraint. Phase optimization for the double-RIS channels is tackled via the alternating optimization and majorization-minimization (MM) schemes. Moreover, closed-form solutions are obtained for each case. Numerical results indicate that the deployment of two relays, one near each RIS, achieves higher rates at low and medium signal-to-noise ratios (SNRs) compared to placing a single relay between the two RISs; while at high SNRs, the latter approach achieves higher rates only if the inter-relay interference for the former case is considerably high. 
	\end{abstract}
	\section{Introduction}
	Conventional active relaying, such as decode-and-forward (DF), is a well known technology that is used to extend the coverage, and/or to enhance the quality-of-service (QoS) between a pair of transceiving nodes \cite{laneman2004cooperative}. Ideally, the locations and/or number of relays should be optimized based on a certain cost function, such as to maximize the rate or to minimize the transmit power while satisfying a given QoS constraint \cite{minelli2014optimal, ngo2011linear}. 
	\par In contrast, the reconfigurable intelligent surface (RIS) technology is a new concept in wireless communications, where a large number of low-cost, nearly-passive reflecting elements are utilized to direct the impinging signal toward a desired destination, such that the multiple signal paths are constructively combined at the receiver \cite{tan2016increasing}. One of the most attractive aspects about RISs is that they do not require power-demanding active radio-frequency chains. Another benefit compared to traditional active relaying is that RISs work on-the-fly, i.e. they do not introduce additional delays due to internal signal processing. Thanks to their low-cost and low power-consumption, it is highly anticipated that RISs will have a key role in future wireless networks \cite{rajatheva2020white, chowdhury20206g, kisseleff2020reconfigurable}. \par However, due to the large degradation of the signal power with distance, which is caused by the absence of active amplification at the RIS and the double path-loss, few studies have shown that considerably large surfaces are required to outperform a conventional single-antenna relay \cite{bjornson2019intelligent, di2020reconfigurable}. Motivated by this fact, many researchers started adopting classical relays, such as DF or amplify-and-forward (AF), to enhance the performance of RIS-assisted networks \cite{abdullah2020hybrid, abdullah2020optimization, wang2021joint, ying2020relay, yildirim2021hybrid, obeed2021joint, huang2021deep, kang2021irs}. However, in all these works, only a single relay was utilized to enhance the RIS-assisted transmission. Moreover, even though the work in \cite{kang2021irs} considers three RISs, only one of them was deployed near the single relay, whereas the other two were located within short distances of the source and destination.
	\par However, in many real-world scenarios, the signal might go through multiple hops before reaching the destination. Therefore, we aim to find the optimal way of combining relays with RISs when there is more than a single RIS between the two transceiving nodes. In particular, we consider the double-RIS reflection case where the signal is subject to reflections from two RISs before reaching the destination, and we propose three different half-duplex (HD) relay-aided network architectures, and compare their effective rates under a total power constraint. The findings of this work can help understand how to perform optimal route optimization, relay placement, and RIS-relay pairing for future multihop RIS-relay assisted networks. 
	\par \textit{Notations}: Matrices and vectors are denoted by boldface uppercase and lowercase letters, respectively.	$\boldsymbol x^T$, $\boldsymbol x^\ast$, and $\left\| \boldsymbol x \right\|$ are the transpose, conjugate, and Euclidean norm of a vector $\boldsymbol x$, respectively, and $[\boldsymbol x]_i$ is the $i$th element of $\boldsymbol x$. $\left|x\right|$ and $\angle{(x)}$ are the absolute value and the phase of a complex number $x$, respectively. $\mathbb E\{a\}$ is the expected value of $a$, while $\boldsymbol I_N$ is the $N\times N$ identity matrix. $\mathrm {diag} \{\boldsymbol x\}$ is a diagonal matrix whose diagonal contains the elements of $\boldsymbol x$, while $\mathrm {diag}\{\boldsymbol X\}$ is a vector whose elements are the diagonal of $\boldsymbol X$. Finally, $\Re{\{x\}}$ denotes the real part of a complex number $x$.
	\section{System Model and Phase Optimization}
	We consider a time division duplex scenario where there is a single-antenna source ($S$) aiming to transmit a signal to a single-antenna destination ($D$) with the help of two RISs ($I_1$ and $I_2$). Due to large distances, obstacles and path-loss, we assume that $S$ has a direct link with only $I_1$, and similarly $D$ has a direct link with only $I_2$.
	\par
\begin{figure}[t]
		\centering
		{\includegraphics[width=8cm,height=9cm, trim={0cm 0cm -0cm 0cm},clip]{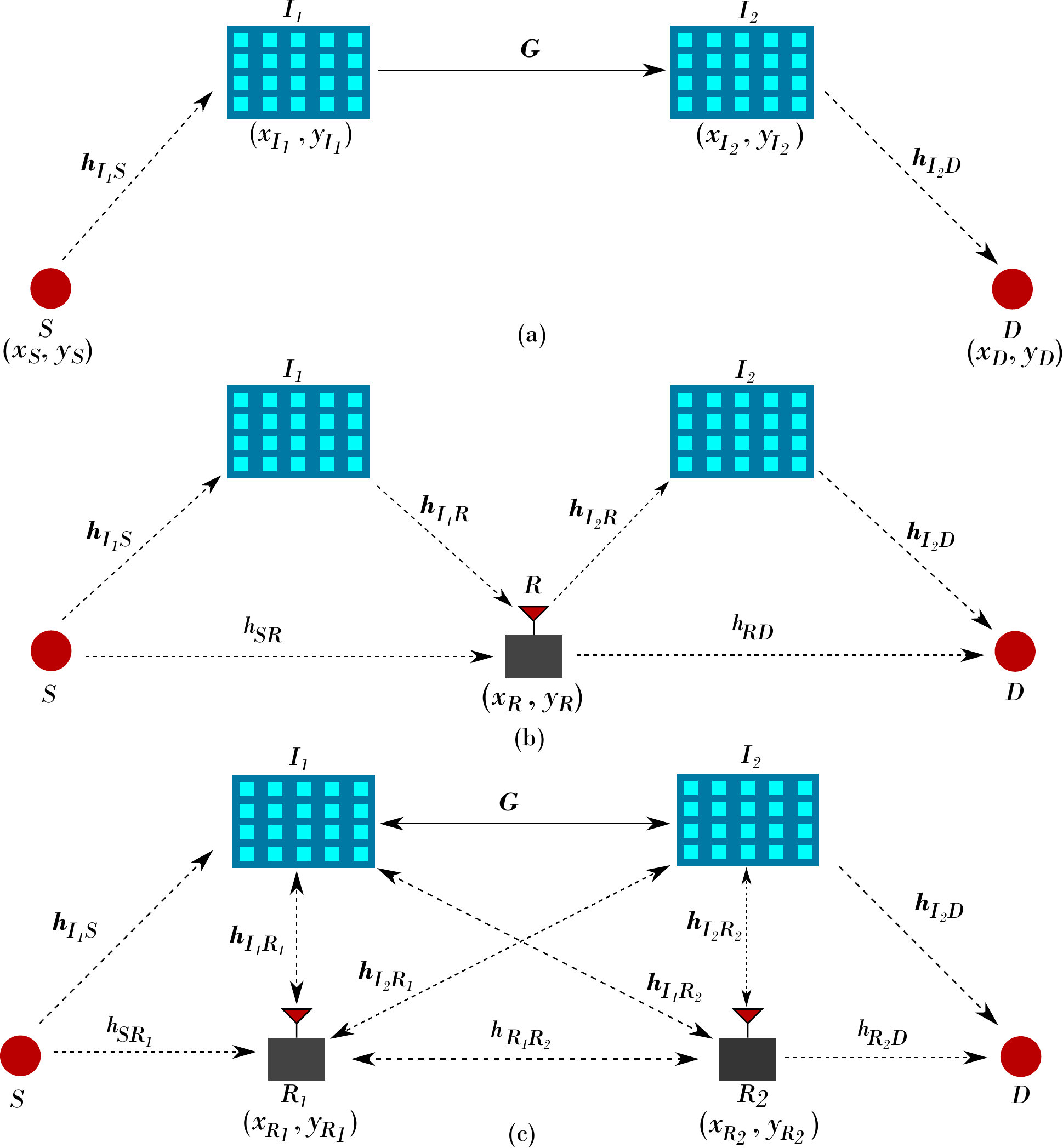}}
		\caption{The considered relay-aided network architectures: (a) Communication through two RISs only, (b) Communication through two RISs and a single relay, and (c) Communication through two RISs and two relays.}
		\label{system}
	\end{figure}

 To enhance the performance of the double-RIS channel, we deploy a single-antenna HD-DF relay(s) between the two ends. In particular, we investigate three different relay-aided scenarios. The first one corresponds to the case of a single relay (R) present between $I_1$ and $I_2$, and the communication takes place over two time-slots. In the second scenario, we assume that there are two relays $R_1$ and $R_2$, placed near $I_1$ and $I_2$, respectively, and only a single node in the network can transmit at any given time-instant. In the last scenario, we aim to enhance the second scenario by allowing the second relay $R_2$ to transmit to $D$ while $S$ transmits its data to $R_1$. Note that throughout this work, we assume perfect channel estimation for all links.\footnote{Note that the authors in \cite{you2020wireless, zheng2021efficient} have proposed channel estimation schemes for the double-RIS channels with satisfactory estimation accuracy.} We also assume centralized processing for the different double-RIS communication schemes. Fig. \ref{system} shows the different relay-aided double-RIS network configurations, where we compare with the no-relay scenario as a benchmark scheme. We next start formulating the received signals and corresponding achievable rates for each scenario. 
	\subsection{Transmission through only the RISs}
	In this scenario, we assume that the transmission is realized through the two RISs only. Therefore, the received signal at the destination can be written as
	\begin{equation}
		\small
		y_D^{(1)}(n) = \sqrt{p} \left(\boldsymbol {h}_{I_2D}^T\boldsymbol {\Phi} \boldsymbol G\boldsymbol \Theta\boldsymbol {h}_{I_1S}\right)x_s(n) + w_D(n),
	\end{equation}where the superscript in $y_D^{(1)}$ indicates that this is a single-hop transmission, $n$ is the time index, $p$ is the total transmit power at any given time-instant $n$, $\boldsymbol {h}_{I_1S} \in \mathbb C^M$ and $\boldsymbol {h}_{I_2D} \in \mathbb C^M$ are the channels between $S\rightarrow I_1$, and $I_2\rightarrow D$, respectively,\footnote{The different narrowband fading channels adopted in this work will be explained in detail in Section \ref{results}.} and $M$ is the number of reflecting elements at each RIS. $\boldsymbol G \in \mathbb C^{M\times M}$ is the channel between the two RISs; while $\boldsymbol \Theta = \mathrm{diag} \{\boldsymbol \theta\}\in \mathbb C^{M\times M}$ and $\boldsymbol \Phi = \mathrm{diag} \{\boldsymbol \phi\} \in \mathbb C^{M\times M}$ are the reflection matrices for $I_1$ and $I_2$, respectively. $x_s$ is the information symbol transmitted from $S$ with $\mathbb E\{|x_s|^2\}=1$, and $w_D \sim \mathcal{CN} (0, \sigma^2)$ is the additive white Gaussian noise (AWGN) at the destination. Therefore, the received signal-to-noise ratio (SNR) at the destination is given as
	{\small	\begin{eqnarray} \label{gamma_d_1}
			\gamma_D^{(1)} & = & \rho \left|\boldsymbol {h}_{I_2D}^T\boldsymbol {\Phi} \boldsymbol G\boldsymbol \Theta\boldsymbol {h}_{I_1S}\right|^2,
	\end{eqnarray}}where $\rho= p/\sigma^2$. The achievable rate in this case is
	\begin{equation}
		\small
		\mathcal R^{(1)} = \log_2\Big(1+\gamma_D^{(1)}\Big).
	\end{equation}Clearly, the achievable rate depends on both $\boldsymbol \Phi$ and $\boldsymbol \Theta$. However, to optimize the achievable rate, we first need to reformulate the cascaded channel as follows:
	{\small \begin{eqnarray}
			\left| \boldsymbol {h}_{I_2D}^T\boldsymbol {\Phi} \boldsymbol G\boldsymbol \Theta\boldsymbol {h}_{I_1S}\right|^2 & = & \left| \boldsymbol \phi^T \text{diag}\{\boldsymbol {h}_{I_2D}\} \boldsymbol G \ \text{diag}\{\boldsymbol {h}_{I_1S}\} \boldsymbol \theta \right|^2 \nonumber \\ & = & \left| \boldsymbol \phi^T \boldsymbol F\boldsymbol \theta\right|^2.
	\end{eqnarray}}Now we can present the following optimization problem
	\begin{subequations} \label{OP1}
		\small
		\begin{align}
			& \hspace{1cm}\underset{\boldsymbol \theta,\ \boldsymbol \phi}{\text{maximize}} \hspace{.3cm} \rho \left| \boldsymbol \phi^T \boldsymbol F\boldsymbol \theta\right|^2 \hspace{4.2cm} (\ref{OP1}) \nonumber \\
			&\hspace{1cm}\text{subject to} \nonumber \\ 
			& \hspace{1cm} \big|[\boldsymbol{\theta}]_m\big| =1, \ \ \ \ \ \big|[\boldsymbol{\phi}]_m\big| =1,  \hspace{0.3cm}\forall m\in\mathcal M,
		\end{align}
	\end{subequations}
	where $\mathcal M = \{1, 2, ..., M\}$ is the set of all reflecting elements at each RIS. The optimization problem in (\ref{OP1}) is non-convex, due to the unit-modulus constraint and the coupled optimization variables. Therefore, we adopt an alternating approach where we fix $\boldsymbol \phi$ to solve for $\boldsymbol \theta$, and vice-versa.
	\subsubsection{Optimizing $\boldsymbol \theta$ for a given $\boldsymbol \phi$}Let $\boldsymbol r^T = \boldsymbol \phi^T \boldsymbol F \in \mathbb C^{1\times M}$, now we can formulate the following optimization problem
	\begin{subequations} \label{OP2}
		\small
		\begin{align}
			& \hspace{2cm} \underset{\boldsymbol \theta }{\text{maximize}} \hspace{.3cm} \rho \left| \boldsymbol r^T \boldsymbol \theta\right|^2  \hspace{3.5 cm } (\ref{OP2}) \nonumber \\
			&\hspace{2cm} \text{subject to}  \hspace{0.5cm}\big|[\boldsymbol{\theta}]_m\big| =1, \hspace{0.3cm}\forall m\in\mathcal M.
		\end{align}
	\end{subequations}
	The solution to the above optimization problem can be given in a closed-form as follows:
	\begin{equation}
		\small
		[\boldsymbol \theta^\star]_m = \exp\Big(j\angle{\big([\boldsymbol r]_m\big)}\Big)^{\ast}.
	\end{equation}
	\subsubsection{Optimizing $\boldsymbol \phi$ for a given $\boldsymbol \theta$}Similar to the previous approach, and after defining $\boldsymbol v = \boldsymbol F \boldsymbol \theta \in\mathbb C^M$, we can write $\big|\boldsymbol \phi^T\boldsymbol F\boldsymbol \theta\big| = \big|\boldsymbol{\phi}^T\boldsymbol v\big|$. Therefore, the solution to $\boldsymbol \phi$ is
	\begin{equation}
		\small
		[\boldsymbol \phi^\star]_m = \exp\Big(j\angle{\big([\boldsymbol v]_m\big)}\Big)^{\ast}.
	\end{equation}We alternate the optimization process between $\boldsymbol \phi$ and $\boldsymbol \theta$ until the increment in the achievable rate between two successive iterations falls below a certain threshold, or we reach a maximum number of optimization iterations. 
	\subsection{Communication through RISs and a single relay}
	In this case, we assume that there is one HD-DF relay ($R$), placed in the middle between $I_1$ and $I_2$,\footnote{The choice of placing the single relay between $I_1$ and $I_2$ comes intuitively to balance the SNRs of first and second hops.} and the transmission is carried out through two time-slots. 
	\subsubsection{First-Hop} In the first time-slot, $S$ transmits its data to $R$ through the direct link and the reflected signal from $I_1$. Therefore, the received signal at $R$ is given as
	\begin{equation}
		\small
		y_R^{(2)}(n) = \sqrt{p}\Big(h_{SR}+\boldsymbol {h}_{I_1R}^T\boldsymbol \Theta \boldsymbol {h}_{I_1S}\Big)x_s(n) + w_R(n),
	\end{equation}
	where the superscript in $y_R^{(2)}$ indicates that there are two hops in this case, $\boldsymbol {h}_{I_1R}\in\mathbb C^M$ is the channel vector between $I_1$ and $R$, $h_{SR}\in \mathbb C$ is the channel between $S$ and $R$, and $w_R\sim \mathcal{CN} (0, \sigma^2)$ is the AWGN at $R$. To maximize the received SNR at $R$,\footnote{Note that there exists another path from $S\rightarrow I_1\rightarrow I_2\rightarrow R$, however, this path is neglected here as the received signal through $S\rightarrow\ I_1\rightarrow R$ will be dominant due to shorter travel distance and less reflections.} the phase-shifts of $\boldsymbol \Theta$ should be selected as follows
	\begin{equation}
		\small
		[\boldsymbol \Theta^\star]_{m,m} = \exp\Big(j\Big(\angle\big(h_{SR}\big) - \angle\big([\boldsymbol {h}_{I_1R}]_{m}[\boldsymbol {h}_{I_1S}]_{m}\big)\Big)\Big),
	\end{equation}$\forall m\in\mathcal M$. Then, the received SNR at $R$ with optimal phase-shifts can be expressed as
	\begin{equation}
		\small
		\gamma_R^{(2)} = \rho \Big(|h_{SR}| + \sum_{m \in \mathcal M} \Big| [\boldsymbol {h}_{I_1R}]_{m}[\boldsymbol {h}_{I_1S}]_{m} \Big|\Big)^2.
	\end{equation}
	\subsubsection{Second-Hop}During the second time-slot, the relay re-transmits the signal, with power $p$, to the destination through the direct link and $I_2$. Assuming successful decoding of $x_s$ at the relay, the received signal at $D$ can be given as 
	\begin{equation}
		\small
		y_D^{(2)}(n+1) = \sqrt{p}\Big(h_{RD}+\boldsymbol {h}_{I_2D}^T\boldsymbol \Phi \boldsymbol {h}_{I_2R}\Big)x_s(n) + w_D(n+1),	
	\end{equation}
	where $h_{RD}\in \mathbb C$ and $\boldsymbol h_{I_2R}\in\mathbb C^M$ are the channels between $R \rightarrow D$ and $R \rightarrow I_2$, respectively, and $w_D\sim \mathcal{CN} (0, \sigma^2)$ is the AWGN at $D$. Assuming perfect phase-shifts at $I_2$ for $\boldsymbol \Phi$, the received SNR at $D$ is
	\begin{equation}
		\small
		\gamma_D^{(2)} = \rho \Big(|h_{RD}| + \sum_{m \in \mathcal M} \Big| [\boldsymbol {h}_{I_2D}]_{m}[\boldsymbol {h}_{I_2R}]_{m} \Big|\Big)^2,
	\end{equation}and the corresponding achievable rate is
	\begin{equation}
		\small
		\mathcal R^{(2)} = \frac{1}{2} \log_2\Big( 1 +\min\{ \gamma_R^{(2)},\  \gamma_D^{(2)} \} \Big),
	\end{equation}and the ($\frac{1}{2}$) pre-log factor is due to the two-hop transmission.
	\subsection{Communication through RISs and two relays}
	In this case, we assume that there are two HD-DF relays, $R_1$ and $R_2$, to assist the communication between $S$ and $D$. In particular, we assume that $R_1$ is placed in close proximity to $I_1$, while $R_2$ is placed near $I_2$. Furthermore, the transmission takes place over three time-slots, since we assume that only one node can transmit at any given time-instant.
	\subsubsection{First-Hop}In the first-hop, $S$ transmits its signal to $R_1$ through the direct link and $I_1$. Therefore, the received signal at $R_1$ is
	\begin{equation}
		\small
		y_{R_1}^{(3,3)}(n) = \sqrt{p}\Big(h_{SR_1}+\boldsymbol {h}_{I_1R_1}^T\boldsymbol \Theta \boldsymbol {h}_{I_1S}\Big)x_s(n) + w_{R_1}(n),
	\end{equation}where the superscript in $y_{R_1}^{(3,3)}$ indicates that there are three hops and $S$ transmits a new block of data every three transmission time-slots, $\boldsymbol h_{I_1R_1}\in\mathbb C^M$ is the channel vector between $I_1$ and $R_1$, and $w_{R_1}\sim \mathcal{CN} (0, \sigma^2)$ is the AWGN at $R_1$. Assuming perfect phase-optimization for $\boldsymbol \Theta$ at $I_1$, the received SNR at $R_1$ is
	\begin{equation}
		\small
		\gamma_{R_1}^{(3,3)} = \rho \Big(|h_{SR_1}| + \sum_{m \in \mathcal M} \Big| [\boldsymbol {h}_{I_1R_1}]_{m}[\boldsymbol {h}_{I_1S}]_{m} \Big|\Big)^2.
	\end{equation}
	\subsubsection{Second-Hop}During the second-hop, $R_1$ re-transmits the signal, assuming successful decoding, to $R_2$ through the direct link, direct reflection links from both $I_1$ and $I_2$, as well as double-reflection link. Therefore, and assuming perfect decoding of $x_s$ at $R_1$, the received signal at $R_2$ can be given as shown in (\ref{y_r2_3}) at the top of the next page, where we have $\boldsymbol h_{I_1R_2} \in\mathbb C^M$, $\boldsymbol h_{I_2R_1}\in\mathbb C^M$, $\boldsymbol h_{I_2R_2}\in\mathbb C^M$ and $h_{R_1R_2} \in \mathbb C$ are the channels between $I_1\rightarrow R_2$, $I_2 \rightarrow R_1$, $I_2 \rightarrow R_2$ and $R_1\rightarrow R_2$, respectively. $\boldsymbol \Psi_1 \in\mathbb C^{M\times M}$ and $\boldsymbol \Psi_2 \in\mathbb C^{M\times M}$ are the reflection matrices for $I_1$ and $I_2$, respectively, during the second-hop, and $w_{R_2} \sim \mathcal{CN} (0, \sigma^2)$ is the AWGN at $R_2$.
	\begin{figure*}[]
		\centering
		{\small \begin{equation} \label{y_r2_3}
				y_{R_2}^{(3,3)}(n+1) = \sqrt{p} \Big( h_{R_1R_2} + \boldsymbol {h}_{I_1R_2}^T \boldsymbol \Psi_1 \boldsymbol{h}_{I_1R_1} + \boldsymbol{h}_{I_2R_2}^T\boldsymbol{\Psi}_2 \boldsymbol{h}_{I_2R_1} + \boldsymbol{h}_{I_2R_2}^T\boldsymbol\Psi_2 \boldsymbol G\boldsymbol \Psi_1 \boldsymbol{h}_{I_1R_1} \Big)x_s(n)+w_{R_2}(n+1).
		\end{equation}}
		\hrule
	\end{figure*}
	\par Let $\boldsymbol Q = \mathrm{diag} \{\boldsymbol {h}_{I_2R_2}\}\boldsymbol G \mathrm{diag}\{\boldsymbol{h}_{I_1R_1}\} \in \mathbb C^{M\times M}$, $\boldsymbol {u}_{I_1} = \mathrm{diag} \{\boldsymbol{h}_{I_1R_2}\} \boldsymbol{h}_{I_1R_1} \in \mathbb C^M$, $\boldsymbol {u}_{I_2} = \mathrm{diag} \{\boldsymbol{h}_{I_2R_2}\} \boldsymbol{h}_{I_2R_1} \in \mathbb C^M$, $\boldsymbol \psi_1=\mathrm{diag}\{\boldsymbol \Psi_1\} \in \mathbb C^M$, and $\boldsymbol \psi_2=\mathrm{diag}\{\boldsymbol \Psi_2\} \in \mathbb C^M$. Then, the received SNR at $R_2$ can be written as follows:
	\begin{equation}
		\small
		\gamma_{R_2}^{(3,3)} = \rho \Big|h_{R_1R_2} + \boldsymbol \psi_1^T \boldsymbol {u}_{I_1} + \boldsymbol\psi_2^T \boldsymbol {u}_{I_2} + \boldsymbol \psi_2^T\boldsymbol Q\boldsymbol \psi_1\Big|^2.
	\end{equation}
	Clearly the SNR depends on both $\boldsymbol{\psi}_1$ and $\boldsymbol \psi_2$, therefore, we can formulate the following optimization problem
	\begin{subequations} \label{OP3}
		\small
		\begin{align}
			& \hspace{0.5cm}\underset{\boldsymbol \psi_1, \ \boldsymbol \psi_2}{\text{maximize}} \hspace{.3cm} \rho \Big|h_{R_1R_2} + \boldsymbol \psi_1^T \boldsymbol {u}_{I_1} + \boldsymbol\psi_2^T \boldsymbol {u}_{I_2} + \boldsymbol \psi_2^T\boldsymbol Q\boldsymbol \psi_1\Big|^2 \hspace*{0.3cm}(\ref{OP3}) \nonumber \\
			& \hspace{0.5cm} \text{subject to} \nonumber \\ 
			& \hspace{0.5cm}\big|[\boldsymbol{\psi}_i]_m\big| =1, \hspace{0.3cm}\forall m\in\mathcal M,\ \ \  i\in{\{1,2\},}
		\end{align}
	\end{subequations}the above optimization problem is also non-convex. Accordingly, we adopt an alternating optimization scheme where we fix one of the optimization variables and solve for the other one. In particular, and for a given $\boldsymbol \psi_2$, we have
	$\gamma_{R_2}^{(3,3)}  = \rho \Big|\boldsymbol z^T \boldsymbol \psi_1 + c\Big|^2$, where $\boldsymbol z^T = \big(\boldsymbol u_{I_1}^T + \boldsymbol \psi_2^T \boldsymbol Q\big)$, and $c = \big(\boldsymbol \psi_2^T \boldsymbol u_{I_2} + h_{R_1R_2}\big)$. Therefore, it is straightforward to see that the optimal phase-shift for the $m$th element of $\boldsymbol \psi_1$ is
	\begin{equation}
		\small
		[\boldsymbol \psi_1^\star]_m = \exp\Big(j\Big(\angle(c) - \angle([\boldsymbol z]_m)\Big)\Big). 
	\end{equation}
	Similarly, to optimize the phase-shifts of $\boldsymbol \psi_2$, we can fix $\boldsymbol \psi_1$ to obtain $\gamma_{R_2}^{(3,3)} = \rho\Big|\boldsymbol v^T\boldsymbol \psi_2 + r\Big|^2$, where $\boldsymbol v = \big(\boldsymbol u_{I_2} + \boldsymbol Q \boldsymbol \psi_1\big)$, and $r = \big(h_{R_1R_2} + \boldsymbol \psi_1^T\boldsymbol u_{I_1}\big)$. It follows that the optimal phase-shift for the $m$th element of $\boldsymbol \psi_2$ is
	\begin{equation}
		\small
		[\boldsymbol \psi_2^\star]_m = \exp\Big(j\Big(\angle(r) - \angle([\boldsymbol v]_m) \Big)\Big). 
	\end{equation}
	\subsubsection{Third-Hop}After receiving the information, $R_2$ will decode the message and then will retransmit it to $D$ through the direct link and $I_2$. Assuming successful decoding at $R_2$, the received signal at the destination can be written as
	\begin{equation}
		\small
		y_D^{(3,3)}(n+2) = \sqrt{p}\Big(h_{R_2D} + \boldsymbol h_{I_2D}^T\boldsymbol \Phi \boldsymbol h_{I_2R_2}\Big)x_s(n)+w_D(n+2),
	\end{equation}
	and the corresponding received SNR at $D$, assuming perfect phase-optimization at $I_2$ for $\boldsymbol \Phi$, can be expressed as
	{\small \begin{align}
			\gamma_D^{(3,3)} = \rho\Big(\big|h_{R_2D}\big| + \sum_{m \in \mathcal M} \Big|[\boldsymbol h_{I_2D}]_m [\boldsymbol h_{I_2R_2}]_m \Big| \Big)^2,
	\end{align}}and the corresponding achievable rate can be written as
	\begin{equation}
		\small
		\mathcal R^{(3,3)} = \frac{1}{3} \log_2\Big(1+\min\{\gamma_{R_1}^{(3,3)}, \gamma_{R_2}^{(3,3)}, \gamma_{D}^{(3,3)}\}\Big),
	\end{equation}where the spectral efficiency is reduced by a factor of $3$ since a new block of data is transmitted every three time-slots. 
	\begin{figure*}[] 
		{\small	\begin{align} \label{y_odd}
				y_{R_1}^{(3,2)}(n_{o}) = & \sqrt{p_1} \underbracket{\Big(h_{SR_1} + \boldsymbol {h}_{I_1R_1}^T\boldsymbol \Theta \boldsymbol {h}_{I_1S} \Big)}_{{h_1}}x_s(n_{o}) + \sqrt{p_2}\underbracket{\Big(h_{R_1R_2} + \boldsymbol{h}_{I_1R_1}^T\boldsymbol \Theta \boldsymbol{h}_{I_1R_2}  + \boldsymbol{h}_{I_2R_1}^T\boldsymbol \Phi \boldsymbol{h}_{I_2R_2}  + \boldsymbol{h}_{I_1R_1}^T\boldsymbol \Theta \boldsymbol G \boldsymbol \Phi \boldsymbol{h}_{I_2R_2} \Big)}_{h_2} x_s(n_{o}-2) \nonumber \\ + & w_{R_1}(n_{o}).
		\end{align}}
		\hrule 
	\end{figure*}
	\begin{figure*}[]
		{\small \begin{equation} \label{canc} 
				\mathbb E\{y_{R_1}^{(3,2)}(n_o) x_s(n_o-2)^*\} = \sqrt{p_1} h_1 \underbracket{\mathbb E\{x_s(n_{o}) x_s(n_o-2)^*\}}_{=0} + \sqrt{p_2} h_2  \underbracket{\mathbb E\{x_s(n_o-2) x_s(n_o-2)^*\}}_{=1} + \underbracket{\mathbb E\{w_{R_1}(n_{o})x_s(n_o-2)^*\}}_{=0} = \sqrt{p_2} {h}_2.
		\end{equation}}
		\hrule
	\end{figure*}
	\subsection{Enhanced transmission with RISs and two-relays}
	The main setback for the previous scenario is the $\frac{1}{3}$ pre-log factor, which can be costly at high SNRs. Therefore, we further present an enhanced transmission scheme such that while $R_2$ is transmitting its signal to $D$, $S$ transmits a new data packet to $R_1$. Note that for the first two time-slots ($n \in \{1,2\}$), all equations in the previous subsection hold in terms of SNRs and phase-optimization; as for the subsequent frames (i.e. when $n>2$), the received signals and transmit powers will change as will be thoroughly explained here. 
	\subsubsection{Received signal at $R_1$}
	At a given odd-time instant $n_{o}$ ($n_{o} \ge 3$), both $S$ and $R_2$ transmit data to $R_1$ and $D$, respectively. Our focus here is on the received signal at $R_1$. 
	\par Clearly, $R_1$ will receive an interfering signal from $R_2$ in addition to the desired signal from $S$, as shown in (\ref{y_odd}) at the top of the next page, where the superscript in $y_{R_1}^{(3,2)}$ denotes that there are $3$ hops, and $S$ transmits a new data packet every two time-slots, $p_1$ and $p_2$ are the transmit powers at $S$ and $R_2$, respectively, with $p_1 + p_2 = p$ to maintain the total transmit power budget, and $\boldsymbol \Theta$ and $\boldsymbol \Phi$ are the reflection matrices at $I_1$ and $I_2$, respectively. The $2$nd term in (\ref{y_odd}) represents the interference from $R_2$, which can be canceled in different ways. For example, if a global channel state information (CSI) is available, then $R_1$ can cancel this interference perfectly (assuming perfect channel estimation), since the signal transmitted from $R_2$, i.e. $x_s(n_o-2)$, can be viewed as the signal that $R_1$ transmitted in the previous time-slot. Otherwise, $R_1$ can estimate the overall effective channel between itself and $R_2$. This can be performed according to the maximum-likelihood estimation by multiplying the received signal at $R_1$ with the conjugate of the transmitted signal from $R_1$ at time $n_o-1$  (which is $x_s(n_o-2)^\ast$), as shown in (\ref{canc}).\footnote{Here we assume that this interference signal is suppressed by one of the two methods explained above. Note that another way of suppressing the interference at $R_1$ is through the passive beamforming at $I_1$ and $I_2$. However, we will leave this approach for investigation in our future work.} After performing interference cancellation at $R_1$, we can rewrite the received signal in (\ref{y_odd}) as follows:
	{\small \begin{align}
			y_{R_1}^{(3,2)}(n_{o}) & = \sqrt{p_1} \Big(h_{SR_1} + \boldsymbol {h}_{I_1R_1}^T\boldsymbol \Theta \boldsymbol {h}_{I_1S} \Big)x_s(n_{o}) \nonumber \\ & + h_{e} x_s(n_o-2) + w_{R_1}(n_o),
	\end{align}}where $h_{e} = \sqrt{p_2}(h_2 - \hat{h}_2)$ is the residual interference cancellation error at $R_1$, which is usually assumed to follow normal distribution such that $h_{e} \sim\mathcal{CN}(0, \sigma_e^2)$. Assuming perfect phase optimization for $\boldsymbol \Theta$ at $I_1$ to maximize the power of received signal at $R_1$ from S, we can formulate the received signal-to-interference plus noise ratio (SINR) at $R_1$ as follows:
	\begin{equation}
		\small
		\gamma_{R_1}^{(3,2)} = \frac{p_1 \Big(|h_{SR_1}| + \sum_{m \in \mathcal M} \Big| [\boldsymbol {h}_{I_1R_1}]_{m}[\boldsymbol {h}_{I_1S}]_{m} \Big|\Big)^2}{\sigma_{e}^2 + \sigma^2}.
	\end{equation}
	Next we focus on the received signal at the destination.\footnote{Note that the received signal at $R_2$ in the next time-slot (i.e. at time ($n_o+1$)) will not be affected by this enhanced transmission scheme, since only $R_1$ will be transmitting data to $R_2$ with a power budget of $p$ while the source will be silent. Therefore, we have $\gamma_{R_2}^{(3,2)} = \gamma_{R_2}^{(3,3)}.$}
	\subsubsection{Received signal at D}
	While $S$ is transmitting its data to $R_1$, $R_2$ transmits the decoded signal from $R_1$ in the previous time-slot to $D$. The received signal at $D$ can be expressed as
	{\small \begin{align}
			y_D^{(3,2)}(n_o) = & \sqrt{p_2} \Big(h_{R_2D} + \boldsymbol {h}_{I_2D}^T \boldsymbol \Phi \boldsymbol {h}_{I_2R_2}\Big) x_s(n_o-2)\nonumber \\ + & \sqrt{p_1}\Big(\boldsymbol{h}_{I_2D}^T\boldsymbol \Phi \boldsymbol {G} \boldsymbol \Theta \boldsymbol {h}_{I_1S} \Big)x_s(n_o) + w_D(n_o).
	\end{align}}Note that $x_s(n_o)$ is intended for $R_1$, and therefore it represents interference to $D$. As a result, the SINR at $D$ is 
	\begin{equation}
		\small
		\gamma_D^{(3,2)} = \frac{p_2\Big|h_{R_2D} + \boldsymbol \phi^T \boldsymbol a\Big|^2}{p_1 \Big|\boldsymbol{\phi}^T \boldsymbol b\Big|^2 + \sigma^2},
	\end{equation}where $\boldsymbol b = \mathrm{diag}\{\boldsymbol h_{I_2D}\}\boldsymbol q$, $\boldsymbol a = \mathrm{diag}\{\boldsymbol h_{I_2D}\}\boldsymbol h_{I_2R_2}$, $\boldsymbol \phi = \mathrm{diag} \{\boldsymbol \Phi\}$, and $\boldsymbol q = \boldsymbol {G} \boldsymbol \Theta \boldsymbol {h}_{I_1S}$.
	Now we can formulate the following optimization problem:
	\begin{subequations} \label{OP4}
		\small
		\begin{align}
			& \hspace{1cm} \underset{\boldsymbol \phi}{\text{minimize}} \hspace{.5cm} u(\boldsymbol \phi) \stackrel{\Delta}{=}\frac{p_1 \Big|\boldsymbol{\phi}^T \boldsymbol b\Big|^2 + \sigma^2}{p_2\Big|h_{R_2D} + \boldsymbol \phi^T \boldsymbol a\Big|^2}  \hspace{1.7cm} (\ref{OP4})\nonumber \\
			&\hspace{1cm} \text{subject to}  \hspace{0.5cm}\big|[\boldsymbol{\phi}]_m\big| =1, \hspace{0.3cm}\forall m\in\mathcal M.
		\end{align}
	\end{subequations}
	This problem belongs to fractional programming \cite{dinkelbach1967nonlinear}. As such, we formulate the following parametric program:
	\begin{subequations} \label{OP5}
		{\small \begin{align}
				& \hspace{0.6cm} \underset{\boldsymbol \phi }{\text{minimize}} \ \ p_1 \Big|\boldsymbol{\phi}^T\boldsymbol b\Big|^2 + \sigma^2-\mu \Big(p_2\Big|h_{R_2D} + \boldsymbol \phi^T \boldsymbol a \Big|^2\Big) \hspace{0.5cm} \nonumber (\ref{OP5}) \\
				&\hspace{0.6cm}\text{subject to}  \hspace{0.5cm}\big|[\boldsymbol{\phi}]_{m}\big| =1, \hspace{0.3cm}\forall m\in\mathcal M.
	\end{align}}\end{subequations}where $\mu\ge0$ is an introduced parameter. Although problem (\ref{OP5}) is non-convex, it can be solved using the iterative majorization-minimization (MM) method. In particular, we can introduce the following upper-bound of (\ref{OP5}) \cite{shen2019secrecy, sun2016majorization}:
	{\small \begin{align} \label{g}
			f(\boldsymbol \phi, \mu) & \stackrel{\Delta}{=}  p_1 \Big|\boldsymbol{\phi}^T\boldsymbol b\Big|^2 + \sigma^2-\mu \Big(p_2\Big|h_{R_2D} + \boldsymbol \phi^T \boldsymbol a \Big|^2\Big) \nonumber \\ & = \boldsymbol \phi^T\boldsymbol X\boldsymbol \phi^\ast - 2\mu p_2\Re\left\{\boldsymbol \phi^T h_{R_2D}^\ast \boldsymbol a\right\}-\mu p_2\left|h_{R_2D}\right|^2 + \sigma^2\nonumber \\ & \le \lambda_{\max}(\boldsymbol X) \left\|\boldsymbol \phi\right\|^2 -2\Re \left\{\boldsymbol \phi^T\boldsymbol \alpha\big(\tilde{\boldsymbol \phi}, \mu \big)\right\} + \beta\big(\tilde{\boldsymbol \phi}, \mu \big),
	\end{align}}where $\boldsymbol X = \big(p_1\boldsymbol b\boldsymbol b^H-\mu p_2\boldsymbol a\boldsymbol a^H\big)$, $\lambda_{\max} (\boldsymbol X)$ is the maximum eigenvalue of $\boldsymbol X$, $\boldsymbol L = \lambda_{\max} (\boldsymbol X)\boldsymbol I_M$, $\boldsymbol \alpha\big(\tilde{\boldsymbol \phi}, \mu\big) = \big(\left(\boldsymbol L-\boldsymbol X \right)\tilde{\boldsymbol \phi}^\ast + \mu p_2 h_{R_2D}^\ast \boldsymbol a \big)$, $\beta\big(\tilde{\boldsymbol \phi}, \mu\big) = \big(\tilde{\boldsymbol \phi}^T\big(\boldsymbol L - \boldsymbol X\big)\tilde{\boldsymbol \phi}^\ast - \mu p_2 \left|h_{R2D}\right|^2 + \sigma^2\big)$, and $\tilde{\boldsymbol \phi}$ is the solution to $\boldsymbol \phi$ in the previous iteration of the MM scheme. Accordingly, minimizing the upper-bound of (\ref{OP5}) can be simplified as:
	\begin{subequations} \label{OP6}
		{\small \begin{align}
				& \hspace{1cm} \underset{\boldsymbol \phi}{\text{minimize}} \hspace{0.5cm} \lambda_{\max}(\boldsymbol X) \left\|\boldsymbol \phi\right\|^2 -2\Re \left\{\boldsymbol \phi^T\boldsymbol \alpha\big(\tilde{\boldsymbol \phi}, \mu \big)\right\} \hspace{0.5cm} \nonumber (\ref{OP6}) \\
				&\hspace{1cm} \text{subject to}  \hspace{0.5cm}\big|[\boldsymbol{\phi}]_{m}\big| =1, \hspace{0.3cm}\forall m\in\mathcal M.
	\end{align}}\end{subequations}For a given value of $\mu$, the term $\lambda_{\max}(\boldsymbol X) \left\|\boldsymbol \phi\right\|^2$ is a constant. As such, the optimal phase for the $m$th element at any given iteration of the MM scheme can be given as follows:
	\begin{equation}\label{phi_opt}
		\small 
		\left[\boldsymbol \phi\right]^\star_m = \exp\Big(j \angle{\big([\boldsymbol \alpha (\tilde{\boldsymbol \phi}, \mu)]_m \big)}\Big)^\ast, 
	\end{equation}and the value of $\mu$ is updated after each iteration as follows:
	\begin{equation}\label{mu} 
		\small
		\mu = \frac{p_1 \Big|(\boldsymbol{\phi}^\star)^T\boldsymbol b\Big|^2 + \sigma^2}{\Big(p_2\Big|h_{R_2D} + (\boldsymbol \phi^\star)^T \boldsymbol a \Big|^2\Big)}.
	\end{equation}However, before the MM algorithm starts, we initialize $\mu$ based on any feasible solution for $\boldsymbol \phi$ in (\ref{mu}), and then utilize both $\mu$ and $\boldsymbol \phi$ to find $\boldsymbol \alpha(\boldsymbol \phi, \mu)$.
	\par \textit{Proposition:} The value of $u(\boldsymbol \phi)$ in (\ref{OP4}) is monotonically non-increasing with $k$, where $k>0$ is the number of iterations for the MM scheme.
	\par \textit{Proof:} see Appendix A. 
	\par For each iteration of the MM scheme, we find the corresponding values of $\mu$ and $\boldsymbol \alpha$, then we optimize the phase-shifts based on (\ref{phi_opt}). The same procedure will be repeated until convergence or reaching a maximum number of iterations. It follows that the achievable rate utilizing this enhanced transmission scheme can be expressed as follows:
	\begin{equation}
		\mathcal{R}^{(3,2)} = \frac{1}{2} \log_2\Big(1 + \min \big\{\gamma_{R_1}^{(3,2)}, \gamma_{R_2}^{(3,2)}, \gamma_{D}^{(3,2)}\big\}\Big).
	\end{equation}
	
	\section{Results and Discussion} \label{results}
	We start by introducing the wireless channels adopted in our work. All links from and to the RISs were assumed to experience Rician fading with both line-of-sight (LoS) and non-LoS (NLoS) channels. In particular, $\boldsymbol G = \sqrt{\frac{K}{K+1}}\boldsymbol G^{\mathrm{LoS}}+\sqrt{\frac{1}{1+K}}\boldsymbol G^{\mathrm{NLoS}}$, where $K$ is the Rician factor, $\boldsymbol G^{\mathrm{LoS}}$ contains the LoS channels, with each link having a deterministic absolute value of $d_{I_1I_2}^{-\tilde{\alpha}/2}$, where $d_{I_1I_2}$ is the distance between two RISs and $\tilde{\alpha}$ is the path-loss exponent for LoS links. In contrast, $\boldsymbol G^{\mathrm{NLoS}}$ is the complex Gaussian Rayleigh fading channel, where each link has a zero mean and a variance of $d_{I_1I_2}^{-\bar{\alpha}}$, where $\bar{\alpha}$ is the path-loss exponent for NLoS channels. Similarly, we have $\boldsymbol h_i = \sqrt{\frac{K}{K+1}}\boldsymbol h_i^{\mathrm{LoS}}+\sqrt{\frac{1}{1+K}}\boldsymbol h_i^{\mathrm{NLoS}}$, where each element of $\boldsymbol h_i^{\mathrm{LoS}}$ has a fixed absolute value of $d_i^{-\tilde{\alpha}/2}$; while $\boldsymbol h_i^{\mathrm{NLoS}} \sim\mathcal{CN} (\boldsymbol 0, \boldsymbol I_M d_i^{-\bar{\alpha}})$, $i\in\{I_1S, I_1R, I_1R_1, I_1R_2, I_2R, I_2R_1, I_2R_2, I_2D\}$. In contrast, we assume pure Rayleigh fading between nodes that do not include any of the two RISs, such that $h_j\sim\mathcal{CN}(0,d_j^{-\bar{\alpha}})$, $j\in\{SR, SR_1, R_1R_2, R_2D\}$. Moreover, $S$ was located at the origin of a $\mathrm{2D}$ plane such that $(x_S, y_S) = (0,0)$, while $(x_{I_1}, y_{I_1}) = (60, 20)$, $(x_{I_2}, y_{I_2}) = (240, 20)$, $(x_{R_1}, y_{R_1})=(60, 0)$, $(x_R, y_R) = (150, 0)$, $(x_{R_2}, y_{R_2}) = (240, 0)$, and $(x_D, y_D) = (300, 0)$, all in meters (see Fig.\ref{system}). In addition, we set $\tilde{\alpha} = 2.3$, $\bar{\alpha} = 3.5$, $\sigma^2=1$, $K=10$ $\mathrm {dB}$; while the maximum number of iterations to optimize any phase-shift vector was $50$ (which was shown to be enough for convergence), and the optimization convergence threshold was set to $10^{-3}$. Furthermore, for the enhanced transmission scheme, we have $p_1 = p_2 = \frac{1}{2}p$. We define the interference-to-noise ratio (INR) as $\rho_e = \sigma_e^2/\sigma^2$, while the transmit SNR was defined as $p/\sigma^2$.
	\begin{figure}[t]
		\centering
		{\includegraphics[width=5.25cm,height=5cm, trim={5cm 8.4cm 5.5cm 8.4cm},clip]{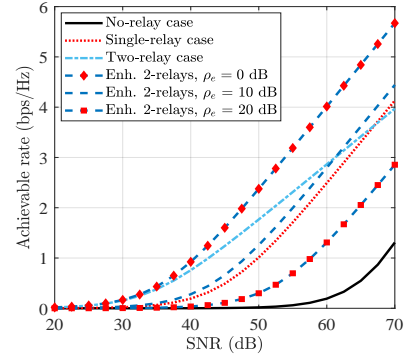}}
		\caption{Achievable rate vs transmit SNR ($\mathrm {dB}$) for different relaying schemes with different levels of INR when $M = 128$.}
		\label{Rate}
	\end{figure}
	\par As demonstrated in Fig. \ref{Rate}, the double-RIS communication without including relaying suffers from a notably low rate performance. This is due to the large loss in signal power due to the lack of active amplification. In contrast, utilizing one or two relays can provide a significant performance gain. To be more specific, and regardless of the value of SNR, adopting the enhanced two-relays transmission is always better than the single-relay case as long as the inter-relay interference is suffiently suppressed, i.e. $\rho_{e} \le 10 \ \mathrm{dB}$. Otherwise, the choice between a single relay and two relays depends purely on the value of SNR. At low SNRs, deploying two relays such that no two nodes can transmit in the network at the same time, can still achieve higher rates than the single relay case despite the $(1/3)$ pre-log penalty; while at high SNRs, the single relay case leads to higher rates. 
	\begin{figure}[t]
		\centering
		{\includegraphics[width=5.25cm,height=5cm, trim={5cm 8.4cm 5.5cm 8.4cm},clip]{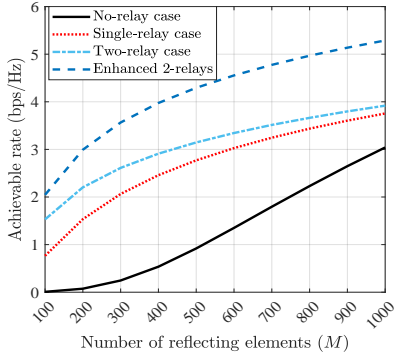}}
		\caption{Achievable rate vs number of reflecting elements per RIS for different relaying schemes when the transmit $\mathrm {SNR} = 50 \ \mathrm{dB}$, and $\rho_e = 0 \ \mathrm{dB}$.}
		\label{differet_M}
	\end{figure}
	\par Finally, Fig. \ref{differet_M} shows the performance of different transmission schemes for a wide range of number of reflecting elements. Once again, our results indicate that adopting two relays to assist the transmission is highly desirable for the double-RIS assisted communication even when the considered RISs are sufficiently large with hundreds of reflecting elements. For example, to achieve $3 \ \mathrm{bps/Hz}$, the enhanced two-relay transmission requires $200$ reflecting elements per RIS, compared to $600$ and $1000$ elements for the single-relay and no-relay cases, respectively, given that the inter-relay interference is suppressed to the noise level.  
	\section{Conclusion and Future work}
	We investigated the DF relay-aided double-RIS reflection channels for coverage extension. Three different relay-aided network architectures were proposed for effective rate maximization under a total power constraint. Our results demonstrated that deploying two relays for the double-RIS channel achieves higher rates at low and medium SNRs; while at high SNRs, deploying a single relay to assist the two RISs is better only if the inter-relay interference was high. The generalization to multihop with arbitrary numbers of RISs and relays is subject to future investigations.  
	\section*{Acknowledgment}
	This work was supported by the Luxembourg National Research Fund (FNR) under the CORE project RISOTTI.
	\section*{Appendix A}
	Let us denote $p_2\Big|h_{R_2D} + \boldsymbol \phi^T \boldsymbol a\Big|^2$ by $f_{a}(\boldsymbol \phi)$, $p_1 \Big|\boldsymbol{\phi}^T \boldsymbol b\Big|^2 + \sigma^2$ by $f_b(\boldsymbol \phi)$, and the right hand side of (\ref{g}) by $g(\boldsymbol \phi|\tilde{\boldsymbol \phi}, \mu)$. Let $\tilde{\boldsymbol \phi}$ and $\boldsymbol \phi^\star$ be the phase-shift values of $I_2$ before and after running a single iteration of the MM scheme, and let $\tilde \mu$ denote the value of $\mu$ that corresponds to $\tilde{\boldsymbol \phi}$. Then, from the left hand side of (\ref{g}), we have $f(\boldsymbol \phi^\star, \tilde{\mu}) = f_b(\boldsymbol \phi^\star) - \tilde{\mu} f_a(\boldsymbol \phi^\star)\stackrel{(\text a)}{\le} g(\boldsymbol \phi^\star|\tilde{\boldsymbol \phi}, \tilde{\mu})\stackrel{(\text b)}{\le}g(\tilde{\boldsymbol \phi}|\tilde{\boldsymbol \phi}, \tilde{\mu}) = f(\tilde{\boldsymbol \phi}, \tilde{\mu}) = f_b(\tilde{\boldsymbol \phi})-\tilde{\mu} f_a(\tilde{\boldsymbol \phi}) \stackrel{(\text c)}{=}0$, where (a) holds from (\ref{g}), (b) holds since $\boldsymbol \phi^\star$ minimizes $g(\boldsymbol \phi|\tilde{\boldsymbol \phi}, \tilde{\mu})$, and (c) holds from the definition of $\mu$ in (\ref{mu}). Therefore, we have $u(\boldsymbol \phi^\star) = f_b(\boldsymbol \phi^\star)/f_a(\boldsymbol \phi^\star) \le \tilde{\mu} = f_b(\tilde{\boldsymbol \phi})/f_a(\tilde{\boldsymbol \phi}) = u(\tilde{\boldsymbol \phi})$.

	\bibliographystyle{IEEEtran}
	\bibliography{Multi_IRSs_relays}	
\end{document}